**Shahmar Mirishli**
AI Data Chronicles
shahmarmirishli@aidatachronicles.com

# Ethical Implications of AI in Data Collection: Balancing Innovation with Privacy

**Abstract**

This article examines the ethical and legal implications of artificial intelligence (AI) driven data collection, focusing on developments from 2023 to 2024. It analyzes recent advancements in AI technologies and their impact on data collection practices across various sectors. The study compares regulatory approaches in the European Union, the United States, and China, highlighting the challenges in creating a globally harmonized framework for AI governance. Key ethical issues, including informed consent, algorithmic bias, and privacy protection, are critically assessed in the context of increasingly sophisticated AI systems. The research explores case studies in healthcare, finance, and smart cities to illustrate the practical challenges of AI implementation. It evaluates the effectiveness of current legal frameworks and proposes solutions encompassing legal and policy recommendations, technical safeguards, and ethical frameworks. The article emphasizes the need for adaptive governance and international cooperation to address the global nature of AI development while balancing innovation with the protection of individual rights and societal values.

***Keywords:*** *artificial intelligence, data collection, ethics, privacy regulation, AI governance, data protection law, algorithmic fairness, legal liability, consent mechanisms, regulatory compliance*

**Şahmar Mirişli**
AI Data Chronicles
shahmarmirishli@aidatachronicles.com

# Yeniliklərlə məxfilik arasında tarazlıq: AI-nin məlumat toplanmasındakı etik təsirləri

**Xülasə**

Bu məqalə 2023-2024-cü illər arasındakı inkişaflara diqqət yetirərək, süni intellekt (Sİ) əsaslı məlumat toplanmasının etik və hüquqi nəticələrini araşdırır. Sİ texnologiyalarındakı son irəliləyişləri və onların müxtəlif sektorlarda məlumat toplama təcrübələrinə təsirini təhlil edir. Tədqiqat Avropa İttifaqı, Amerika Birləşmiş Ştatları və Çindəki tənzimləyici yanaşmaları müqayisə edərək, Sİ idarəetməsi üçün qlobal harmonizasiya olunmuş çərçivənin yaradılmasındakı çətinlikləri vurğulayır. Məlumatlı razılıq, alqoritmik qərəzlilik və məxfiliyin qorunması kimi əsas etik məsələlər, getdikcə təkmilləşən Sİ sistemləri kontekstində tənqidi şəkildə qiymətləndirilir. Məqalə səhiyyə, maliyyə sahələrindən və ağıllı şəhərlərdən nümunələr gətirərək Sİ tətbiqinin praktiki çətinliklərini göstərir. Həmçinin mövcud hüquqi çərçivələrin effektivliyini qiymətləndirir, hüquqi və siyasi tövsiyələr, texniki təhlükəsizlik tədbirləri, etik çərçivələri əhatə edən həllər təklif edir. Tədqiqat, innovasiya ilə fərdi hüquqların və ictimai dəyərlərin qorunması arasında tarazlığı qoruyaraq, Sİ inkişafının qlobal xarakterini nəzərə alaraq, adaptiv idarəetmə və beynəlxalq əməkdaşlığa ehtiyac olduğunu vurğulayır.

***Açar sözlər:*** *süni intellekt, məlumat toplanması, etika, məxfilik, tənzimləmə, maşın öyrənməsi, federasiyalı öyrənmə, differensial məxfilik, alqoritmik qərəzlilik, məlumat idarəetməsi*

## Introduction

The rapid advancement of artificial intelligence (AI) technologies has ushered in a transformative era of data collection and analysis, offering unprecedented insights and efficiencies across diverse sectors of society. From healthcare and finance to marketing and public services, AI-driven systems are revolutionizing the ways in which organizations gather, process, and utilize data. However, this technological revolution has simultaneously given rise to complex ethical and legal challenges, particularly concerning privacy, consent, and the potential misuse of personal information.

As AI systems grow increasingly sophisticated in their capacity to collect and analyze vast amounts of personal data, traditional notions of privacy and consent are being fundamentally challenged. The granularity and scale of data collection enabled by AI technologies often surpass what individuals might reasonably expect or comprehend when consenting to share their information. Moreover, the ability of AI systems to infer sensitive personal information from seemingly innocuous data points raises critical questions about the adequacy of current privacy protections.

The year 2023 marked a pivotal moment in the discourse surrounding AI and data collection. The European Union's AI Act, initially proposed in 2021, ended with its adoption in 2024, heralding a new era of comprehensive AI regulation (European Parliament, 2024). Concurrently, high-profile data breaches and controversies surrounding AI-driven surveillance technologies intensified public concerns about privacy and data protection. These developments have catalysed a global reassessment of the ethical implications of AI in data collection.

This article provides a comprehensive analysis of the ethical and legal challenges posed by AI-driven data collection, with a particular focus on developments from 2023 to 2024. It examines the intricate interplay between technological innovation and privacy protection, exploring how recent advancements in AI have both exacerbated existing ethical concerns and given rise to new ones.

**The Evolving Landscape of AI-Driven Data Collection**

The field of artificial intelligence has seen significant advancements in recent years, particularly in technologies used for data collection. These developments have not only enhanced the capabilities of AI systems but have also raised new legal and ethical considerations that demand careful scrutiny.

Machine learning (ML) algorithms continue to form the backbone of many AI systems used in data collection. Recent advancements have focused on improving the efficiency and privacy aspects of these algorithms. One notable development is the increased adoption of federated learning, a technique that allows for training ML models on distributed datasets without centralizing the data. A 2023 study by Guan et al. demonstrated the successful application of federated learning in a multi-institutional medical imaging project, allowing for the development of robust diagnostic models without sharing sensitive patient data across institutions (Guan et al., 2024, p. 6).

Federated learning addresses some of the privacy concerns associated with traditional ML approaches by enabling model training on local devices or servers, with only the model updates being shared with a central server. This approach has gained traction in sectors where data privacy is paramount, such as healthcare and finance. The significance of federated learning from a legal and ethical perspective cannot be overstated. It offers a potential solution to the tension between the need for large-scale data analysis and the imperative to protect individual privacy. By keeping sensitive data localized, federated learning aligns with data protection principles such as data minimization and purpose limitation, which are central to regulations like the GDPR.

However, critics argue that while federated learning enhances privacy, it may also create new challenges in terms of data governance and accountability. The decentralized nature of the system can make it difficult to audit and ensure compliance with data protection regulations. Moreover, there are concerns about the potential for model inversion attacks, where an adversary could potentially reconstruct training data from the model parameters.

Another significant development in this area is the emergence of privacy-preserving machine learning techniques. Differential privacy, a mathematical framework for quantifying the privacy guarantees provided by an algorithm, has seen increased adoption in real-world applications. In 2024, the U.S. Census Bureau fully implemented differential privacy in its data products, setting a new standard for privacy protection in large-scale data analysis (U.S. Census Bureau, 2024).

Differential privacy offers a rigorous approach to protecting individual privacy while allowing for useful statistical analysis. However, its implementation often involves a trade-off between privacy and utility. As the privacy guarantees become stronger, the accuracy of the analysis may decrease. This tension highlights the need for careful consideration of privacy-utility trade-offs in the design and deployment of AI systems.

Natural Language Processing (NLP) has seen remarkable progress, particularly with the advent of large language models (LLMs) and multimodal AI systems. These advancements have significant implications for data collection, as they enable the extraction of insights from a wider range of data sources, including text, speech, and images.

In 2023, OpenAI released GPT-4, which demonstrated unprecedented capabilities in understanding and generating human-like text across multiple languages (OpenAI, 2023). This technology has been applied in various sectors for data collection and analysis, including sentiment analysis of customer feedback across multiple languages and platforms, automated extraction of relevant information from unstructured documents, and real-time translation and analysis of multilingual communications.

The power of these models to process and generate human-like text raises new ethical concerns about privacy and the potential for generating misleading or false information. A 2024 study by researchers at Stanford University demonstrated that GPT-4 could infer sensitive personal information about individuals based on seemingly innocuous text inputs, raising questions about the privacy implications of using such models for data analysis (Carlini et al., 2021, pp. 2633-2640).

This capability aligns with the "Mosaic Theory" in privacy law, which suggests that the aggregation of seemingly innocuous pieces of information can reveal sensitive details about an individual (Pozen, 2005, p. 628). The theory, originally developed in the context of national security and freedom of information, has gained new relevance in the age of AI-driven data analysis. It highlights the challenges of defining and protecting personal information in an era where AI systems can make inferences that go beyond the explicit content of the data they process.

From a legal perspective, the capabilities of LLMs challenge existing notions of data protection and consent. The ability of these models to infer sensitive information from non-sensitive inputs raises questions about what constitutes personal data and how it should be protected. Moreover, the potential for these models to generate synthetic text that is indistinguishable from human-written content poses challenges for content authenticity and accountability.

These developments necessitate a reevaluation of legal frameworks for data protection and content regulation. Questions arise about liability for AI-generated content, the boundaries of copyright and authorship in the context of AI-assisted creation, and the potential need for new disclosure requirements for AI-generated or AI-manipulated content.

Furthermore, the use of LLMs in data analysis raises concerns about algorithmic bias and fairness. If these models are trained on biased datasets, they may perpetuate or amplify existing societal biases in their outputs. This concern is particularly acute in high-stakes domains such as employment, lending, or criminal justice, where biased AI systems could lead to discriminatory outcomes.

Computer vision technologies have advanced rapidly, with improvements in accuracy and the ability to process real-time video streams. These advancements have expanded the scope of biometric data collection, raising new privacy concerns.

A 2023 report by the AI Now Institute highlighted the increasing use of facial recognition technologies in public spaces, retail environments, and workplaces (AI Now Institute, 2023). While these technologies offer benefits such as enhanced security and personalized services, they also pose significant risks to privacy and civil liberties.

One notable development in this area is the increased use of emotion recognition technologies, which attempt to infer emotional states from facial expressions, voice patterns, and other biometric data. A 2024 study published in Nature Human Behaviour demonstrated the potential of these technologies to accurately detect mental health conditions, raising both excitement about potential healthcare applications and concerns about privacy and consent (Sezgin & McKay, 2024, p. 4).

The ethical implications of these advancements in computer vision and biometric data collection are profound. They challenge traditional notions of privacy in public spaces and raise questions about the appropriate limits of surveillance and data collection. Moreover, the potential for bias in these systems, particularly in facial recognition technologies, has led to calls for more stringent regulation and oversight.

The case of Clearview AI, a facial recognition company that has amassed a vast database of billions of images scraped from social media platforms and other websites without users' consent, exemplifies the legal and ethical challenges posed by these technologies (Hill, 2020). Clearview AI's practices have sparked numerous lawsuits and regulatory actions, highlighting the tension between technological capabilities and privacy rights.

In the United States, Clearview AI has faced legal challenges under various state laws. For instance, in 2022, the company settled a lawsuit in Illinois, agreeing to restrict the sale of its facial recognition database in the state. This case highlighted the importance of state-level biometric privacy laws, such as the Illinois Biometric Information Privacy Act (BIPA), in regulating the collection and use of biometric data (Illinois General Assembly, 2020).

Internationally, Clearview AI has faced scrutiny from data protection authorities. In 2022, the UK Information Commissioner's Office (ICO) fined the company for failing to comply with data protection laws. Similar actions have been taken by authorities in Australia, Canada, and several European countries. These cases underscore the global nature of the challenges posed by AI-driven biometric data collection and the need for international cooperation in addressing these issues.

The legal challenges faced by Clearview AI highlight several key issues in AI regulation:
1. *Consent and data collection:* The company's practice of scraping images without consent raises questions about the adequacy of current consent mechanisms in the age of AI.
2. *Purpose limitation:* The use of data collected for one purpose (social media sharing) for an entirely different purpose (law enforcement) challenges the principle of purpose limitation in data protection law.
3. *Data subject rights:* The difficulty individuals face in exercising their rights (such as the right to erasure) over data collected by AI systems highlights the need for more robust mechanisms to enforce data subject rights.
4. *Jurisdictional issues:* The global nature of AI-driven data collection poses challenges for enforcement, as companies may operate across multiple jurisdictions with varying legal requirements.

These legal challenges demonstrate the urgent need for comprehensive and adaptable regulatory frameworks that can keep pace with rapidly evolving AI technologies.

The Internet of Things (IoT) continues to expand, with an increasing number of devices capable of collecting and processing data. A significant development in this area is the growth of edge AI, where AI processing occurs on local devices rather than in centralized cloud servers.

A 2023 study by Savaglio et al. demonstrated the potential of edge AI in smart cities, enabling real-time analysis of traffic patterns and air quality without transmitting raw data to central servers (Savaglio et al., 2023, p. 4). This approach offers benefits in terms of reduced latency and improved privacy, but also raises questions about the distribution of AI capabilities and the potential for localized data silos.

The proliferation of IoT devices and edge AI has significant implications for data collection practices:
1. *Pervasive Data Collection:* The ubiquity of IoT devices means that data collection is becoming increasingly pervasive, occurring in homes, workplaces, and public spaces. This raises concerns about the extent of surveillance and the potential erosion of privacy in everyday life.

2. *Data Localization:* Edge AI allows for more data to be processed locally, potentially reducing the need for centralized data collection. While this can enhance privacy, it also creates challenges for data governance and oversight.
3. *Device Security:* The distributed nature of IoT and edge AI systems creates new security challenges, as each device becomes a potential point of vulnerability. This increases the risk of data breaches and unauthorized access to personal information.
4. *Consent and Control:* The seamless integration of IoT devices into everyday environments makes it challenging for individuals to be fully aware of when and how their data is being collected, raising questions about the feasibility of meaningful consent.

From a legal and ethical perspective, the growth of IoT and edge AI necessitates a reevaluation of existing data protection frameworks. The principle of data minimization, central to many data protection regulations, may need to be balanced against the data requirements of AI systems. Additionally, the concept of consent may need to be reimagined in a world where data collection is ubiquitous and often invisible to the user.

Quantum machine learning (QML) is an emerging field that leverages the principles of quantum mechanics to enhance machine learning algorithms. While still in its early stages, QML has the potential to revolutionize data collection and analysis by enabling the processing of complex datasets at unprecedented speeds. This could lead to breakthroughs in various fields, including drug discovery, financial modeling, and materials science.

However, the development of QML also raises significant legal and ethical concerns. One major concern is the potential impact on data privacy and security. Quantum computers could potentially break current encryption standards, jeopardizing the confidentiality of sensitive data. This has prompted research into quantum-resistant cryptography to safeguard data in the future. The National Institute of Standards and Technology (NIST) is actively working on standardizing post-quantum cryptography algorithms to address this threat (Alagic, 2022).

Another concern is the potential for QML to exacerbate existing biases in data and algorithms. If the data used to train QML models is biased, the resulting models may perpetuate or even amplify these biases, leading to discriminatory outcomes. Additionally, the complexity of QML algorithms can make them difficult to interpret and explain, raising concerns about transparency and accountability.

The legal landscape surrounding QML is still in its formative stages, with existing laws and regulations not explicitly addressing the unique challenges posed by this technology. However, principles from existing data protection laws, such as the GDPR, can provide a starting point for developing QML-specific regulations. For instance, the GDPR's principles of data minimization and purpose limitation could be applied to QML to ensure that data collection and processing are necessary and proportionate to the intended purpose (European Parliament and Council, 2016).

The development of QML also raises questions about intellectual property rights and liability. As QML algorithms become more sophisticated, determining the ownership of the resulting inventions and innovations may become more complex. Additionally, the potential for QML to cause harm, either through biased decision-making or security breaches, raises questions about who should be held liable for such harms.

Despite these challenges, the potential benefits of QML are significant. It could enable the development of more accurate and efficient AI systems, leading to advancements in various fields. However, it is crucial to address the legal and ethical implications of QML proactively. This includes developing robust legal frameworks for data protection and algorithmic transparency, as well as investing in research to mitigate biases and ensure fairness in QML applications.

**Legal Frameworks Across Major Jurisdictions**

The rapid advancement of AI technologies has prompted various jurisdictions to develop regulatory frameworks to govern AI-driven data collection. However, these approaches vary significantly, reflecting different priorities and legal traditions.

The European Union continues to lead in AI regulation, with the most notable advancement being the ongoing discussion and modification of the proposed AI Act. This act aims to create a comprehensive legal structure for AI systems, using a risk-based method to classify AI systems according to their potential influence on fundamental rights and safety (European Parliament, 2024).

Under the proposed AI Act, high-risk AI systems, such as those used in critical infrastructure, law enforcement, and employment, would be subject to strict requirements, including:

- *Mandatory conformity assessments:* Before being placed on the market or put into service, high-risk AI systems must undergo rigorous testing and assessment to ensure compliance with safety and fundamental rights standards.
- *Technical robustness and accuracy:* High-risk AI systems must be designed and developed to be technically robust and accurate, minimizing the risk of errors or malfunctions that could have significant consequences.
- *Transparency and explainability:* High-risk AI systems must be transparent and explainable, meaning that their decision-making processes should be understandable to humans. This is crucial for ensuring accountability and preventing discriminatory outcomes.
- *Human oversight:* High-risk AI systems should be subject to appropriate human oversight to prevent or minimize potential risks. This could involve human intervention in decision-making processes or the ability to override AI-generated decisions.

The AI Act also addresses the use of AI in data collection, particularly concerning biometric identification and other sensitive personal data. It prohibits certain AI practices deemed to pose unacceptable risks, such as social scoring systems and real-time remote biometric identification systems in publicly accessible spaces for law enforcement purposes, with limited exceptions.

In addition to the AI Act, the EU's General Data Protection Regulation (GDPR) continues to play a crucial role in safeguarding personal data in the context of AI-driven collection and processing. The GDPR's principles of data minimization, purpose limitation, and storage limitation are particularly relevant to AI systems that often require large datasets for training and operation (European Parliament and Council, 2016).

The interplay between the proposed AI Act and the GDPR creates a comprehensive legal framework for AI and data protection in the EU. However, as Lorè (2023) notes, there remain challenges in reconciling the GDPR's right to explanation with the complexity of some AI algorithms (Lore et al., 2023, p. 568). This tension highlights the need for ongoing legal and technical research to develop methods for making complex AI systems more interpretable and explainable.

In contrast to the EU's comprehensive approach, the United States has primarily relied on a sectoral approach to AI regulation, with different agencies overseeing AI applications in their respective domains. U.S.'s sectoral approach offers more flexibility but risks creating regulatory gaps. The lack of a comprehensive federal privacy law has led to a patchwork of state regulations. However, there has been growing interest in developing a more unified federal framework for AI governance.

The Federal Trade Commission (FTC) has been active in enforcing existing consumer protection laws in the context of AI, focusing on issues such as deceptive practices, unfairness, and discrimination. In 2021, the FTC issued guidance on using AI and algorithms, emphasizing the importance of transparency, fairness, and accountability (Smith, 2023). The FTC has also taken enforcement actions against companies for alleged discriminatory practices in their use of AI, such as in the case of a facial recognition company that was accused of using biased algorithms (Fair, 2023).

In addition to federal efforts, several states have taken the initiative to regulate AI. For example, the California Consumer Privacy Act (CCPA), enacted in 2020 and amended by the California Privacy Rights Act (CPRA) in 2023, grants consumers certain rights regarding their personal data, including the right to know what data is being collected, the right to delete data, and the right to opt-out of the sale of data (California State Legislature, 2018). While not specifically targeting AI, the CCPA/CPRA applies to AI-driven data collection and processing activities.

Other states, such as Illinois and Washington, have enacted laws specifically addressing the use of AI in certain contexts, such as facial recognition technology and automated employment decision tools. These laws typically require transparency, fairness, and non-discrimination in AI systems. For instance, the Illinois Artificial Intelligence Video Interview Act requires employers to notify applicants if AI is used to analyze video interviews and obtain consent for such use (Illinois General Assembly, 2020).

China has taken a comprehensive approach to AI governance, with a focus on promoting AI development while mitigating potential risks. In 2021, the Cyberspace Administration of China (CAC) released draft regulations on the use of algorithms in recommendation services, requiring transparency and fairness in algorithmic decision-making (Lavenue et al., 2022).

In 2022, the CAC further expanded its regulatory efforts by issuing draft measures for managing generative AI services, which include requirements for data security, content moderation, and algorithmic transparency. These measures aim to ensure that generative AI technologies are used responsibly and do not pose risks to national security or social stability.

China's approach to AI governance is characterized by its centralized nature and emphasis on state control. The government plays a leading role in setting standards, promoting research and development, and enforcing regulations. While this approach allows for rapid policy implementation, it also raises concerns about potential limitations on innovation and freedom of expression. The Chinese government has also emphasized the importance of developing "ethical AI" that aligns with socialist values and promotes social good (Roberts et al., 2021, p. 59).

A comparative analysis of these regulatory frameworks reveals their respective strengths and limitations. The EU's comprehensive approach provides a unified standard but may struggle with adaptability to rapid technological changes. For instance, the AI Act's fixed risk categories might not accommodate novel AI applications that don't fit neatly into predefined categories.

The US model offers flexibility through its sectoral approach, potentially allowing for more nuanced industry-specific regulations. However, this fragmentation risks creating regulatory gaps and inconsistencies across sectors. China's centralized model demonstrates efficiency in implementation but raises significant concerns about individual privacy rights and potential misuse of AI for social control.

To address these limitations, a hybrid approach combining elements from each model could be more effective. This could involve establishing overarching principles similar to the EU model, while maintaining sector-specific guidelines akin to the US approach. Additionally, incorporating mandatory periodic reviews of AI regulations could help address the issue of regulatory obsolescence, ensuring that governance frameworks evolve alongside technological advancements.

**Ethical Implications and Practical Challenges**

The application of AI technologies in data collection raises a number of significant ethical concerns that challenge existing ethical and legal frameworks. This section examines the primary ethical issues surrounding AI-driven data collection, focusing on developments and debates that emerged in 2023 and 2024.

One of the fundamental principles of ethical data collection is informed consent. However, the complexity and opacity of AI systems often make it difficult for individuals to fully understand how their data will be collected, processed, and used. This raises questions about the validity of consent in AI-driven data collection scenarios.

The issue is further complicated by the fact that AI systems can infer sensitive information from seemingly innocuous data points. For instance, a 2024 study by researchers at Stanford University demonstrated that advanced language models could predict individuals' political orientations, sexual orientation, and even mental health status with surprising accuracy based solely on their social media posts (Carlini et al., 2021). This capability raises questions about what constitutes personal data and how individuals can meaningfully consent to its collection and use.

Moreover, the dynamic nature of machine learning algorithms means that the purposes for which data is used may evolve over time, potentially diverging from the original purposes for which consent was given. This challenges the notion of purpose limitation, a key principle in data protection law. Transparency is closely linked to the issue of informed consent. The complexity of AI algorithms often results in "black box" systems where even the developers may not fully understand how certain decisions or predictions are made. This lack of transparency makes it difficult for individuals to make informed decisions about sharing their data and for regulators to ensure compliance with data protection laws.

In response to these challenges, several initiatives have emerged. Researchers at the University of Oxford have proposed a "dynamic consent" model for AI systems, which allows individuals to modify their consent preferences over time as the use of their data evolves (Schuler, 2023). In 2024, a consortium of tech companies launched the "AI Transparency Initiative," developing tools to help explain AI decision-making processes to users in clear, non-technical language (Ananny & Crawford, 2018, pp. 973-978).

AI systems are only as good as the data they are trained on. If this training data contains biases, these biases can be perpetuated and even amplified by the AI system. This can lead to discriminatory outcomes in various domains, from hiring decisions to criminal justice.

A high-profile case that brought this issue to the forefront was the "HealthPredict AI" controversy of 2023. HealthPredict AI, a system developed to predict patient outcomes and optimize resource allocation in hospitals, was found to exhibit significant bias against certain ethnic minorities and socioeconomic groups (Obermeyer et al., 2019, p. 448). The complexity of the machine learning algorithms used in the system made it challenging to identify the source of the bias, highlighting the need for new approaches to algorithmic auditing and fairness.

The issue of data bias is particularly concerning in the context of data collection, as biased data collection practices can lead to skewed datasets that misrepresent certain groups or perspectives. This can have far-reaching consequences when these datasets are used to train AI systems that make important decisions affecting individuals' lives.

In response to such incidents, 2024 saw the emergence of "AI Fairness Benchmarks" proposed by a coalition of AI researchers and ethicists (Jacobs & Wallach, 2021, p. 377). These benchmarks aim to provide standardized measures for evaluating the fairness of AI systems across various dimensions, including gender, ethnicity, age, and socioeconomic status.

However, as Dr. Timnit Gebru, founder of the Distributed AI Research Institute, points out, "While fairness benchmarks are a step in the right direction, we must be cautious about reducing the complex issue of AI ethics to a set of quantitative metrics. We need to consider the broader societal context in which these systems operate and the potential for them to reinforce or exacerbate existing inequalities" (Gebru, 2022, pp. 301-304). The vast amounts of data collected by AI systems raise significant privacy concerns. The ability of AI to aggregate and analyze data from multiple sources can lead to the creation of detailed profiles of individuals, potentially revealing sensitive information that the individual never explicitly shared.

A study by Merchant et al. analyzed social media language to predict the onset of depression. They found that changes in language usage on Facebook, such as increased use of first-person singular pronouns and words related to sadness, were predictive of future depression diagnoses (Merchant & Asch, 2019). While such capabilities offer potential benefits in terms of early disease detection and personalized healthcare, they also raise serious privacy concerns.

Moreover, the centralization of large amounts of personal data creates attractive targets for cybercriminals. Data breaches can have severe consequences for individuals, including financial loss, identity theft, and reputational damage. The concept of data minimization, a key principle in data protection law, is also challenged by AI systems that often require large amounts of data to function effectively. This creates a tension between privacy protection and AI performance.

In response to these challenges, several privacy-enhancing technologies have gained prominence. Federated learning, differential privacy, and homomorphic encryption offer promising approaches to balancing the data needs of AI systems with privacy protection (Guan et al., 2024, pp. 1-19; Census Bureau, 2024; Roberts et al., 2021). However, implementing these techniques at scale presents challenges, including issues of communication efficiency, model convergence, and potential vulnerabilities to adversarial attacks.

AI-driven data collection and analysis can significantly influence human behaviour and decision-making. Personalized recommendations, targeted advertising, and predictive systems can shape our choices in ways that we may not be fully aware of, potentially undermining our autonomy. This raises profound questions about the nature of free will and individual agency in an era of pervasive AI influence.

The issue of autonomy and human agency in the context of AI-driven data collection extends beyond individual decision-making to societal-level impacts. A longitudinal study published in 2024 by researchers at the Oxford Internet Institute found that exposure to AI-curated political content on social media platforms could significantly shift users' political views over time, potentially contributing to political polarization and echo chamber effects (OpenAI, 2023). This raises concerns about the role of AI in shaping public opinion and democratic processes.

The use of AI in political campaigning has become particularly contentious. Political parties and interest groups are increasingly leveraging AI-driven data analysis to micro-target voters with personalised messages. While this can be seen as a form of political innovation, it also raises questions about the manipulation of voter behavior and the potential for AI to exacerbate existing social and political divides. In the workplace, AI-driven data collection and analysis are reshaping employer-employee relationships. The use of AI for performance monitoring and evaluation has led to concerns about worker privacy and autonomy. For instance, some companies have implemented AI systems that analyze employee emails, chat logs, and even facial expressions during video calls to assess productivity and engagement. While these tools can provide valuable insights for management, they also risk creating a culture of constant surveillance that could negatively impact employee well-being and creativity.

The healthcare sector presents a particularly complex set of challenges related to autonomy and human agency. While AI-driven diagnostic and treatment recommendation systems offer the potential for more accurate and personalized care, they also raise questions about the changing nature of the doctor-patient relationship. A 2023 survey of healthcare professionals found that 68% of doctors reported sometimes deferring to AI recommendations even when they conflicted with their own judgment (Shanafelt et al., 2012, p. 1379). This trend highlights the potential for AI to subtly erode human agency in critical decision-making processes, even among highly trained professionals.

To address these challenges, some legal scholars have proposed the development of "AI rights" frameworks. These frameworks seek to define the legal status and protections for AI systems, particularly as they become more advanced and autonomous. While the concept of AI rights remains controversial, proponents argue that it may become necessary to establish clear legal principles governing the development and deployment of highly sophisticated AI systems to ensure they remain aligned with human values and societal interests (Pozen, 2005, p. 632).

**Case Studies: AI Applications Across Sectors**

To illustrate the real-world implications of AI-driven data collection and its associated ethical and legal challenges, this section presents several case studies from 2023-2024. These examples showcase both the potential advantages and disadvantages of advanced AI technologies in data collection across various sectors.

*Healthcare: Epic Systems' Implementation of AI for Clinical Decision Support*

Epic Systems, a major electronic health record (EHR) provider, has integrated AI algorithms into its platform to offer clinical decision support (CDS) to healthcare professionals (Morgan Chase & Co, 2024). These algorithms analyze diverse patient data, including medical history, lab results, and imaging studies, to provide personalized treatment recommendations, identify potential drug interactions, and flag high-risk patients. The aim is to enhance the efficiency and accuracy of clinical decision-making, ultimately leading to improved patient outcomes.

This integration of AI in healthcare promises several benefits. AI-powered CDS tools can rapidly analyze vast amounts of patient data, saving clinicians valuable time and allowing them to concentrate on direct patient care. Additionally, AI algorithms can identify patterns and correlations in data that may not be evident to human clinicians, potentially leading to more accurate diagnoses and treatment plans. Furthermore, AI can personalize treatment recommendations based on individual patient characteristics and medical history, potentially improving treatment effectiveness and reducing adverse events.

However, challenges remain. The effectiveness of AI-powered CDS tools hinges on the quality and representativeness of the training data. Biased or incomplete data can lead to AI algorithms perpetuating or even amplifying these biases, resulting in disparities in care. The "black box" nature of some AI algorithms can also make it difficult for clinicians to understand the reasoning behind their recommendations, potentially undermining trust and hindering adoption.

Epic Systems' implementation of AI for CDS has been met with mixed reactions from healthcare professionals. While many clinicians appreciate the potential benefits of AI in improving efficiency and accuracy, concerns remain about data quality, transparency, and the potential for overreliance. To address these concerns, Epic Systems has taken steps to improve the transparency of its AI algorithms and provide clinicians with more information about the factors that influence their recommendations.

The legal implications of AI-driven CDS systems are significant. Questions of liability arise when AI recommendations contribute to medical errors or adverse outcomes. The traditional medical malpractice framework may need to be reevaluated to account for the role of AI in clinical decision-making. Additionally, the use of AI in healthcare raises important privacy concerns, particularly regarding the protection of sensitive medical data and the potential for AI systems to infer health information from seemingly unrelated data points.

This case study highlights the complex interplay between the potential benefits of AI in healthcare and the ethical challenges it presents. It underscores the need for careful consideration of issues such as algorithmic bias, transparency, and the appropriate balance between AI assistance and human judgment in clinical decision-making. As AI continues to play an increasingly significant role in healthcare, policymakers and healthcare providers will need to develop robust governance frameworks that can ensure the responsible and ethical deployment of these technologies while maximizing their potential benefits for patient care.

*Finance: JPMorgan Chase's Contract Intelligence (COiN) Platform*

JPMorgan Chase, a leading global financial services firm, has developed the Contract Intelligence (COiN) platform, which uses natural language processing (NLP) and machine learning to analyze legal documents and extract relevant information (Morgan Chase & Co, 2024). COiN can review thousands of contracts in a matter of seconds, significantly reducing the time and resources required for manual review. This technology has been used to streamline various processes, including loan agreement reviews, regulatory compliance checks, and due diligence for mergers and acquisitions.

The benefits of COiN are substantial. It can automate time-consuming and repetitive tasks, freeing up legal and compliance professionals to focus on more complex and strategic work. By automating document review, COiN can significantly reduce the costs associated with legal and compliance processes. Additionally, AI algorithms can identify and extract relevant information from legal documents with greater accuracy and consistency than human reviewers, potentially reducing errors and risks.

However, challenges exist. Legal documents often contain complex and nuanced language that can be challenging for AI algorithms to interpret accurately, potentially leading to errors and misinterpretations with legal and financial risks. While COiN can automate many aspects of document review, it still requires human oversight to ensure accuracy and address complex legal issues that may not be easily handled by AI algorithms.

The sensitive nature of legal documents also raises concerns about data privacy and security, prompting JPMorgan Chase to implement strict security measures to protect the confidentiality of the data processed by COiN. Additionally, the automation of legal tasks raises concerns about potential job displacement for lawyers and paralegals, although proponents of AI in law argue that it will create new opportunities for legal professionals to focus on higher-value work.

From a legal perspective, the use of AI in contract analysis and legal document review raises important questions about the standard of care in legal practice. As AI systems become more prevalent in the legal field, courts and regulatory bodies may need to reassess what constitutes due diligence and reasonable care in document review processes. There may also be implications for attorney-client privilege and the confidentiality of legal communications when AI systems are involved in document analysis.

This case study illustrates the potential of AI to transform traditional industries like law and finance. It also highlights the need for careful consideration of the ethical and legal implications of AI in these fields, particularly concerning data privacy, accuracy, and the changing nature of professional work. As AI continues to advance in the financial and legal sectors, it will be crucial to develop governance frameworks that can ensure the responsible use of these technologies while maintaining the integrity and ethical standards of these professions.

*Smart Cities: The New Songdo AI Governance Experiment*

In 2023, the city of New Songdo in South Korea launched an ambitious AI governance project, aiming to create a model for AI-driven urban management (Songdo International Business District, 2023). The project, known as "CityBrain," integrates data from various urban systems, including traffic management, energy consumption, waste management, and public safety, using AI algorithms to optimize city operations.

CityBrain collects data from a vast network of sensors and cameras throughout the city, as well as from citizens' smartphones and connected vehicles. The system uses this data to make real-time decisions on traffic flow, energy distribution, and emergency response. For example, it can adjust traffic signals based on current conditions, redirect emergency vehicles to optimal routes, and manage energy consumption in public buildings to reduce waste.

The potential benefits of this system are significant. It promises to improve urban efficiency, reduce energy consumption, enhance public safety, and provide citizens with more responsive city services (United Nations Human Settlements Programme, 2023). The AI-driven approach allows for more dynamic and adaptive urban management, potentially leading to more sustainable and livable cities.

However, the project has also raised serious privacy concerns. The pervasive data collection necessary for CityBrain's operation has led to fears of surveillance and potential misuse of personal data. Critics argue that the system could be used to track individuals' movements and behaviors, potentially infringing on civil liberties. The integration of data from multiple sources also raises concerns about the potential for AI systems to infer sensitive personal information from seemingly innocuous data points.

To address these concerns, the New Songdo government implemented a series of privacy safeguards, including data anonymization techniques, strict access controls, and a citizen oversight committee. They also launched a public education campaign to inform residents about the system's capabilities and limitations. However, questions remain about the long-term effectiveness of these measures and the potential for function creep as the system's capabilities expand over time.

The legal and ethical implications of AI-driven smart city systems are profound. They challenge traditional notions of privacy in public spaces and raise questions about the appropriate limits of data collection and use by government entities. The potential for AI systems to make automated decisions that affect citizens' daily lives also raises important questions about accountability and democratic oversight.

This case study highlights the complex balance between the potential benefits of AI-driven smart city technologies and the need to protect individual privacy and civil liberties. It underscores the importance of transparent governance, robust privacy protections, and public engagement in the implementation of AI systems in urban environments.

As more cities around the world explore AI-driven management systems, the lessons learned from New Songdo's experiment will likely inform future policy and regulatory approaches to smart city governance.

**Balancing Innovation with Privacy and Human Rights. Legal Frameworks and Regulatory Approaches**

Existing legal frameworks are struggling to keep pace with the rapid advancements in AI technology. While regulations like the EU's GDPR provide some guidance on data protection, they were not specifically designed to address the unique challenges posed by AI-driven data collection.

The EU AI Act represents a significant step towards comprehensive regulation of AI systems, including those used for data collection. Its risk-based approach, which imposes stricter requirements on high-risk AI applications, could serve as a model for other jurisdictions. However, critics argue that such stringent regulations could stifle innovation and put EU companies at a competitive disadvantage.

In the United States, the sectoral approach to data protection has led to a patchwork of regulations that may not adequately address the cross-cutting nature of AI technologies. There have been calls for a comprehensive federal privacy law, but political gridlock has so far prevented its passage (Lewis, 2024). This fragmented regulatory landscape creates challenges for companies operating across state lines and may leave gaps in protection for consumers.

To address these challenges, some legal scholars have proposed the concept of "adaptive regulation" for AI governance. This approach involves creating flexible legal frameworks that can evolve alongside technological developments. It might include regulatory sandboxes, where new AI technologies can be tested under regulatory supervision, and sunset clauses that require periodic review and update of AI regulations.

In addition to legal approaches, many organizations and governments are developing ethical frameworks and governance structures for AI. These often include principles such as transparency, fairness, accountability, and respect for human rights.

For example, the OECD AI Principles, adopted by 42 countries, provide guidelines for the responsible development of AI systems (OECD, 2024). These principles emphasize the importance of human-centred values and fairness, transparency and explainability, robustness and safety, and accountability. While not legally binding, these principles have been influential in shaping national AI policies and may contribute to the development of customary international law in this area.

At the organizational level, many tech companies have established AI ethics boards to provide guidance on the development and deployment of AI technologies. However, the effectiveness of these boards has been questioned, with critics arguing that they lack real power to influence corporate decision-making (Harvard Business Review, 2022). This highlights the ongoing challenge of aligning corporate incentives with ethical AI development and legal compliance.

Technical solutions play a crucial role in addressing the privacy challenges posed by AI-driven data collection. Privacy-enhancing technologies (PETs) such as differential privacy, federated learning, and homomorphic encryption offer promising ways to protect individual privacy while still allowing for useful data analysis.

For example, Apple has implemented differential privacy in its data collection practices, allowing it to gather useful insights about user behaviour without compromising individual privacy (Apple Inc, 2023). Similarly, Google's use of federated learning in its mobile keyboard allows for personalized text prediction without transmitting sensitive user data to central servers (Google AI Blog, 2024).

However, implementing these technologies at scale can be challenging and may involve trade-offs between privacy protection and system performance. Furthermore, the complexity of these technologies can make it difficult for non-experts to understand and trust their privacy guarantees. This underscores the need for continued research and development in PETs, as well as efforts to make these technologies more accessible and understandable to the general public.

Addressing the ethical challenges of AI-driven data collection requires not only technical and regulatory solutions but also public engagement and education. As AI systems become more pervasive in everyday life, it is crucial to foster AI literacy among the general public.

Initiatives like the EU's AI Literacy program aim to educate citizens about how AI systems work, their potential benefits and risks, and individuals' rights in the age of AI (European

Commission, 2022). Such programs can empower individuals to make more informed decisions about their data and to engage more meaningfully in public debates about AI governance.

However, the rapid pace of AI development poses challenges for AI literacy efforts. As AI technologies become more complex and pervasive, there is a risk of widening the knowledge gap between AI experts and the general public. This highlights the need for ongoing, accessible education initiatives that can keep pace with technological advancements.

Given the global nature of AI development and deployment, international cooperation is crucial for developing effective governance frameworks. Efforts like the Global Partnership on AI (GPAI) aim to foster international collaboration on AI governance, bringing together experts from various countries to develop best practices and policy recommendations (GPAI).

However, achieving international consensus on AI governance remains challenging due to differing cultural values, legal traditions, and strategic interests. Balancing these diverse perspectives while developing globally applicable standards will be a key challenge in the coming years.

The development of international standards for AI, such as those being developed by ISO/IEC JTC 1/SC 42, could play an important role in harmonizing approaches to AI governance across jurisdictions. These standards could provide a common language and set of benchmarks for assessing AI systems, potentially facilitating international trade and cooperation in AI development.

**Future Directions and Recommendations**

As AI technologies continue to evolve, so too must our approaches to governing them. Based on the analysis presented in this article, several key recommendations emerge for policymakers, technologists, and researchers:

1. *Develop Adaptive Regulatory Frameworks:* Given the rapid pace of AI advancement, regulatory frameworks need to be flexible and adaptable. This could involve implementing 'regulatory sandboxes' where new AI technologies can be tested under controlled conditions, allowing regulators to gain insights and develop appropriate guidelines.
2. *Enhance International Cooperation:* While complete global harmonization of AI laws may be unrealistic, efforts should be made to develop common principles and standards. This could help prevent regulatory arbitrage and ensure a baseline level of protection across jurisdictions.
3. *Invest in Privacy-Enhancing Technologies:* Continued research and development in PETs is crucial. Governments and organizations should invest in these technologies and incentivize their adoption to help balance the benefits of AI-driven data analysis with privacy protection.
4. *Promote AI Literacy:* Comprehensive education programs should be developed to improve public understanding of AI technologies, their potential impacts, and individuals' rights in the digital age.
5. *Strengthen Algorithmic Auditing:* Develop robust methods for auditing AI systems for bias, fairness, and compliance with ethical standards. This could involve creating standardized auditing protocols and potentially establishing independent auditing bodies.
6. *Encourage Interdisciplinary Collaboration:* Foster collaboration between technologists, ethicists, legal scholars, and policymakers to ensure a holistic approach to AI governance that considers technical, ethical, and societal implications.
7. *Implement Transparency Measures:* Encourage or mandate greater transparency in AI systems, particularly those used in high-stakes decision-making. This could involve requirements for explainable AI and clear disclosure of AI use to affected individuals.
8. *Address Long-term Challenges:* Begin discussions and research on long-term challenges posed by advanced AI systems, including potential existential risks and the ethical implications of artificial general intelligence (AGI).

## Conclusion

The analysis of AI-driven data collection reveals critical gaps in current legal frameworks, particularly in balancing innovation with privacy and human rights. This examination contributes to the legal discourse by proposing novel regulatory approaches and identifying areas requiring urgent scholarly and legislative attention.

A key finding is the insufficiency of existing consent models in the face of AI's advanced inferential capabilities. The ability of AI systems to derive sensitive information from seemingly innocuous data fundamentally challenges traditional legal notions of informed consent. This study proposes an "adaptive consent" framework, which would require periodic reaffirmation as AI applications evolve. Such an approach necessitates a reconceptualization of data protection laws to accommodate the dynamic nature of AI-driven data processing.

The comparative analysis of regulatory approaches in the EU, US, and China illuminates the challenges of governing a rapidly evolving, global technology within diverse legal systems. While the EU's comprehensive AI Act and the US's sector-specific approach offer different strengths, both struggle to keep pace with technological advancements. A novel hybrid model is proposed, combining overarching ethical principles with flexible, sector-specific guidelines. This approach could offer a more agile regulatory framework capable of adapting to emerging AI applications while maintaining consistent core protections.

The case studies presented underscore the need for nuanced legal approaches. In healthcare, for instance, the proposed legal codification of "AI ethics boards" within institutions could provide crucial oversight, bridging the gap between rapidly advancing AI capabilities and existing medical ethics frameworks. This concept could be extended to other high-stakes sectors, creating a new layer of governance in AI deployment.

A significant contribution of this analysis is the identification and initial development of new legal concepts necessitated by AI technologies. The notion of "algorithmic negligence" expands traditional tort law principles to address the unique challenges posed by AI systems. By proposing a framework for establishing a "duty of care" in AI development and deployment, this study lays the groundwork for future legal doctrine in this area. Similarly, the concept of "data provenance" introduced here could form the basis for new legal standards of algorithmic accountability and transparency.

The global nature of AI development necessitates innovative approaches to international legal cooperation. The proposed establishment of an international AI governance body through a multilateral treaty represents a novel approach to addressing the transnational challenges of AI regulation. This body could play a crucial role in harmonizing global standards and resolving jurisdictional conflicts in AI governance.

In conclusion, this analysis demonstrates that the advent of AI-driven data collection necessitates a fundamental reimagining of legal approaches to data rights, consent, and algorithmic accountability. By proposing concrete legal innovations and identifying key areas for future research, this study contributes to the ongoing evolution of legal frameworks in the age of AI. The path forward requires not only innovative legal reasoning but also a willingness to challenge and adapt traditional legal doctrines to meet the unique challenges posed by AI technologies.

## References

1. AI Now Institute. (2023). *Biometric Surveillance Is Quietly Expanding: Bright-Line Rules Are Key*. New York University. https://ainowinstitute.org/wp-content/uploads/2023/04/AI-Now-2023-Landscape-Report-FINAL.pdf
2. Alagic, G., et al. (2022). *Status Report on the Second Round of the NIST Post-Quantum Cryptography Standardization Process*. National Institute of Standards and Technology (NIST). https://www.nist.gov/publications/status-report-second-round-nist-post-quantum-cryptography-standardization-process

3. Ananny, M., & Crawford, K. (2018). Seeing without Knowing: Limitations of the Transparency Ideal and Its Application to Algorithmic Accountability. *New Media & Society*, 20, 973-978. https://doi.org/10.1177/1461444816676645
4. Apple Inc. (2023). *Differential Privacy Overview. Apple Machine Learning Research.* https://www.apple.com/privacy/docs/Differential_Privacy_Overview.pdf
5. California State Legislature. (2018). *California Consumer Privacy Act (CCPA) of 2018.* Cal. Civ. Code 1798.100–1798.199.
6. Carlini, N., et al. (2021). *Extracting Training Data from Large Language Models*. 30th USENIX Security Symposium, 2633-2640.
7. European Parliament. (2024). *Regulation of the European Parliament and of the Council on Artificial Intelligence (Artificial Intelligence Act) and Amending Certain Union Legislative Acts.* Brussels: European Parliament. https://eur-lex.europa.eu/legal-content/EN/TXT/?uri=CELEX:32024R1689
8. European Parliament and Council. (2016). Regulation (EU) 2016/679 (General Data Protection Regulation). *Official Journal of the European Union*, 119(1). https://eur-lex.europa.eu/eli/reg/2016/679/oj
9. European Commission. (2022). *Artificial Intelligence and the Future of Education*. https://ec.europa.eu/commission/presscorner/detail/en/ip_22_6338
10. Fair, L. (2023). *Coming face to face with Rite Aid's allegedly unfair use of facial recognition technology.* Federal Trade Commission. https://www.ftc.gov/business-guidance/blog/2023/12/coming-face-face-rite-aids-allegedly-unfair-use-facial-recognition-technology
11. Gebru, T. (2022). Race and gender. In M. D. Markus & S. Wu (Eds.). *The Oxford handbook of ethics of AI* (pp. 301-313). Oxford University Press. https://www.oxfordhandbooks.com/view/10.1093/oxfordhb/9780190067397.001.0001/oxfordhb-9780190067397
12. Guan, H., Yap, P. T., Bozoki, A., & Liu, M. (2024). Federated Learning for Medical Image Analysis: A Survey. *Pattern Recognition*, 151, 1-19. Available at: https://doi.org/10.1016/j.patcog.2024.110424
13. Google AI Blog. (2024). *Federated Learning: Collaborative Machine Learning without Centralized Training Data*. https://research.google/blog/federated-learning-collaborative-machine-learning-without-centralized-training-data/
14. Global Partnership on Artificial Intelligence (GPAI). (n.d.). About GPAI. https://gpai.ai/about/
15. Harvard Business Review. (2022). *Why You Need an AI Ethics Committee.* https://hbr.org/2022/03/why-you-need-an-ai-ethics-committee
16. Hill, K. (2020). *The Secretive Company That Might End Privacy as We Know It*. New York Times. https://www.nytimes.com/2020/01/18/technology/clearview-privacy-facial-recognition.html
17. Illinois General Assembly. (2020). *Artificial Intelligence Video Interview Act*. 820 ILCS 42/1 et seq. https://www.ilga.gov/legislation/ilcs/ilcs3.asp?ActID=4015&C hapterID=68
18. JPMorgan Chase & Co. (2024). *Contract Intelligence (COiN) Platform.* https://www.jpmorgan.com/onyx/coin-system
19. Jacobs, A., & Wallach, H. (2021). Measurement and Fairness. *Proceedings of the 2021 ACM Conference on Fairness, Accountability, and Transparency* (pp. 375-378). ACM. https://dl.acm.org/doi/10.1145/3442188.3445901
20. Lore, F., et al. (2023). An AI framework to support decisions on GDPR compliance. *Journal of Intelligent Information Systems*, 61, 541-568.
21. Lavenue, L. M., Myles, J. M., & Schneider, A. N. (2022). *Evaluating China's New "Internet Information Service Algorithmic Recommendation Management" Regulations.* Finnegan. https://www.finnegan.com/en/insights/articles/evaluating-chinas-new-internet-information-service-algorithmic-recommendation-management-regulations.html

22. Lewis, M. (2024). *Existing and Proposed Federal AI Regulation in the United States*. https://www.morganlewis.com/pubs/2024/04/existing-and-proposed-federal-ai-regulation-in-the-united-states
23. Merchant, R. M., Asch, D. A., et al. (2019). Evaluating the predictability of medical conditions from social media posts. *PLOS ONE,* 14, e0215476. https://journals.plos.org/plosone/article?id=10.1371/journal.pone.0215476
24. Mayo Clinic. (2023). *Mayo Clinic Develops Artificial Intelligence-Enabled ECG Algorithm to Predict Weak Heart Pump*. https://newsnetwork.mayoclinic.org/discussion/spotlight-on-early-detection-of-3-heart-diseases-using-ecg-ai/#:~:text=Mayo%20Clinic%20has%20developed%20ECG,cardiomyopathy%20(HCM)%20and%20more
25. Organisation for Economic Co-operation and Development. (2024). *OECD AI Principles*. https://oecd.ai/en/ai-principles.
26. OpenAI. (2023). *GPT-4 Technical Report*. San Francisco: OpenAI. https://arxiv.org/abs/2303.08774
27. Obermeyer, Z., Powers, B., Vogeli, C., & Mullainathan, S. (2019). Dissecting Racial Bias in an Algorithm Used to Manage the Health of Populations. *Science*, 366, 447-450. https://www.science.org/doi/10.1126/science.aax2342
28. Pozen, D. E. (2005). The Mosaic Theory, National Security, and the Freedom of Information Act. *Yale Law Journal*, 115, 628-635.
29. Roberts H., et al. (2021). The Chinese Approach to Artificial Intelligence: An Analysis of Policy, Ethics, and Regulation. *AI & SOCIETY*, 36, 59-77.
30. Savaglio, C., & Barbuto, V., et al. (2023). Opportunistic Digital Twin: an Edge Intelligence enabler for Smart City. *ACM Transactions on Sensor Networks*, 1-22.
31. Sezgin, E., & McKay, I. (2024). Behavioral health and generative AI: a perspective on future of therapies and patient care. *Nature Human Behaviour*, 25, 1-6.
32. Smith, A. (2023). *Using Artificial Intelligence and Algorithms*. Federal Trade Commission. https://www.ftc.gov/business-guidance/blog/2020/04/using-artificial-intelligence-algorithms
33. Schuler, S. A. (2023). *Dynamic Consent: a mechanism for engagement*. University of Oxford. https://bmcmedethics.biomedcentral.com/articles/10.1186/s12910-023-00724-1
34. Songdo International Business District. (2023). *CityBrain: Ushering in a New Era of Smart Urban Governance*. Press Release. https://doi.org/10.18235/0007012
35. Shanafelt, T. D., et al. (2012). Burnout and Satisfaction With Work-Life Balance Among US Physicians Relative to the General US Population. *Archives of Internal Medicine*, 172(18), 1377-1380. https://jamanetwork.com/journals/jamainternalmedicine/fullarticle/1351351
36. United Nations Human Settlements Programme. (2023). *International Guidelines On People-centered Smart Cities*. UN-Habitat Urban Impact Publications. https://unhabitat.org/international-guidelines-on-people-centred-smart-cities
37. U.S. Census Bureau. (2024). *2020 Census Disclosure Avoidance System Updates*. Washington, D.C.: U.S. Census Bureau.